%% file: main.tex
\documentclass[11pt,a4paper]{article}

\usepackage[margin=1in]{geometry}
\usepackage{times}
\usepackage[T1]{fontenc}
\usepackage[utf8]{inputenc}
\usepackage{hyperref}
\usepackage{graphicx}
\usepackage{amsmath,amssymb}
\usepackage{booktabs}
\usepackage{float}
\usepackage{tabularx}
\usepackage{caption}
\usepackage{enumitem}
\usepackage{multirow}
\usepackage{xcolor}
\usepackage{natbib}
\usepackage{url}
\usepackage{microtype}
\usepackage{algorithm}
\usepackage{algpseudocode}

\hypersetup{
    colorlinks=true,
    linkcolor=blue,
    citecolor=blue,
    urlcolor=blue
}

\title{A Statistical-Finance Benchmark for Same-Day Directional Stock Prediction:\\Walk-Forward Evidence from SPY}

\author{Alex Chen\\
  Software Engineer (Independent Researcher)\\
  Previously: Department of Computer Science, University of Montreal\\
  \texttt{alex.chen.1@umontreal.ca}
}

\date{}

\begin{document}
\maketitle

\begin{abstract}
We study statistical predictability in daily U.S. equity prices using only information available at the market open.
Using the SPDR S\&P 500 ETF (SPY) from February 1, 1993 through March 15, 2024 (7,837 trading days), we benchmark XGBoost against Random Forest, LightGBM, Logistic Regression, and naive baselines under expanding-window walk-forward validation.
The prediction task is deliberately modest: after observing the current day's opening price and two lagged target-specific prices, predict whether the same day's close will be above or below the previous day's close.

On the last 800 trading days, Logistic Regression attains the highest close-direction accuracy (71.09\%), while Random Forest is the strongest tree baseline (61.20\%) and XGBoost reaches 58.45\% with a 95\% bootstrap confidence interval of [54.94\%, 62.08].
XGBoost remains useful because it produces continuous price forecasts that support threshold-conditioned analysis: for the close target, directional accuracy rises from 58.4\% on the full test set to 72.7\% when the predicted move exceeds 1\%, but the usable sample falls from 799 observations to 154.
We also report Diebold--Mariano and McNemar tests, regime-specific results for crisis, bull, and COVID-era subsamples, and SHAP feature importance for interpretability.

To place the SPY case study in context, we summarize an auxiliary stock-universe screen covering 541 U.S. equities evaluated with the same XGBoost threshold-AUC metric.
That screen shows substantial heterogeneity across assets and targets; notably, SPY ranks 4th of 541 for the close target, which supports its use as a close-price benchmark while cautioning against broad generalization.
The evidence supports a narrow statistical-finance conclusion: simple daily equity features contain detectable same-day directional information, but the result should be interpreted with careful sample-size accounting and limited economic claims.

\end{abstract}

\section{Introduction}

Forecasting daily stock direction from information available at the open is a natural statistical-finance problem.
It is also a good stress test for simple machine learning models: the signal is weak, market conditions change over time, and even small forms of information leakage can produce overstated results.
These features make careful experimental design at least as important as model choice.
The resulting question sits between market-efficiency tests and empirical machine learning for asset prices \citep{fama1970efficient, lo2004adaptive, gu2020empirical}.

This paper studies a narrow and reproducible variant of the problem.
After observing day-$t$ opening price information, we predict whether the same day's target price will finish above or below the previous day's target price.
Our main inferential focus is the close target on SPY, a liquid and diversified U.S. equity ETF with a long daily history.
We also report auxiliary results for high and low targets, because all three targets are informative in the underlying empirical setup.

The study is motivated by three practical considerations.
First, threshold-conditioned directional accuracy is more informative than raw regression error alone for this forecasting problem.
Second, a focused SPY benchmark can be complemented by a broader stock-universe screen, allowing us to distinguish instrument-specific findings from broader descriptive patterns.
Third, strong economic claims require execution detail beyond what is available in this study, so the paper emphasizes statistical benchmarking rather than a complete trading-system evaluation.

Our main contributions are:
\begin{enumerate}[noitemsep]
    \item We provide a clean walk-forward benchmark on SPY close prediction using XGBoost, Random Forest, LightGBM, Logistic Regression, and naive directional baselines, with statistical significance tests and bootstrap confidence intervals.
    \item We report threshold-conditioned directional accuracy with full sample-size disclosure, showing how apparent high-accuracy regions depend on rapidly shrinking support.
    \item We present regime-specific results across the 2007--2012 crisis period, the 2013--2019 bull market, and the 2020--2023 COVID-era period.
    \item We summarize an auxiliary 541-equity stock screen to contextualize SPY and to show that target-specific predictability is highly heterogeneous across assets.
    \item We use SHAP values to describe which lagged features dominate the XGBoost close-target forecasts.
    \item We separate the statistical forecasting evidence from trading-system claims and document the stock-only reproducibility path.
\end{enumerate}

The remainder of the paper is organized as follows.
Section~\ref{sec:related} reviews related work.
Section~\ref{sec:design} describes the data and experimental design.
Section~\ref{sec:results} presents the empirical results.
Section~\ref{sec:discussion} discusses limitations and interpretation.
Section~\ref{sec:reproducibility} describes reproducibility and scope.
Section~\ref{sec:conclusion} concludes.

\section{Related Work}
\label{sec:related}

\subsection{Machine Learning for Financial Prediction}

Stock prediction has been studied with both classical time-series models and modern machine learning methods.
ARIMA and related linear models remain useful baselines but struggle with the nonlinear and regime-dependent structure of financial data \citep{ariyo2014stock, cavalcante2016computational}.
Neural sequence models such as LSTMs and CNNs have therefore become common in the literature \citep{selvin2017stock, fischer2018deep, hoseinzade2019cnnpred}.
Transformer-style models have also been proposed for financial series \citep{liu2019transformer, ding2020hierarchical}.

\subsection{Tree Ensembles and Financial Tabular Data}

Tree-based ensembles remain strong competitors in finance, especially on small or medium tabular datasets.
XGBoost \citep{chen2016xgboost} and LightGBM \citep{ke2017lightgbm} are attractive because they model nonlinear interactions without requiring extensive normalization or architecture tuning.
\citet{krauss2017deep} found that tree ensembles remained competitive with deep learning in equity-statistical-arbitrage settings, while \citet{borisov2022deep} argued more broadly that tree models remain highly effective on tabular problems.
For financial surveys, see \citet{henrique2019literature} and \citet{sezer2020financial}.

\subsection{Directional Prediction}

Directional prediction is often more relevant than exact price prediction for decision-making.
\citet{leung2000forecasting} emphasized that directional measures can be more appropriate than pure regression metrics in financial settings.
\citet{nyberg2011forecasting} studied directional stock-market forecasting with dynamic binary probit models, providing a classical econometric reference point for direct classification.
\citet{ballings2015evaluating} compared multiple machine learning classifiers for directional stock prediction and found strong performance from tree-based methods.
This paper follows that literature but evaluates direction in a same-day open-to-close/high/low setting using a strict walk-forward protocol.

\subsection{Positioning of This Paper}

This paper is deliberately narrower and more conservative than a full trading-system study.
It evaluates a small family of models under a transparent validation design, emphasizes confidence-threshold trade-offs, and treats the broader stock-universe screen as descriptive context rather than as the source of primary inference.
This distinction is important because financial prediction results are vulnerable to overfitting, multiple testing, and overly optimistic backtest interpretation \citep{harvey2015backtesting, bailey2014probability, lopez2018advances}.

\section{Data and Experimental Design}
\label{sec:design}

\subsection{SPY Dataset}

Our primary dataset is daily SPY OHLC data from Yahoo Finance, spanning February 1, 1993 through March 15, 2024.
The file contains 7,837 trading days, each with opening, high, low, close, and volume fields.
SPY is an attractive benchmark because it is liquid, diversified, and widely used in empirical finance as a proxy for large-cap U.S. equity exposure.
All results in this paper are based on equity data only.
Cryptocurrency data and cryptocurrency-specific training routines are outside the empirical scope.

\subsection{Prediction Setup}

For a target price series $p_2 \in \{\text{high}, \text{low}, \text{close}\}$ and opening price series $p_1 = \text{open}$, we construct the day-$t$ feature vector
\begin{equation}
\mathbf{x}^{(t)} = \big[p_1^{(t-2)},\; p_2^{(t-2)},\; p_1^{(t-1)},\; p_2^{(t-1)},\; p_1^{(t)}\big].
\label{eq:features}
\end{equation}
The target is the same-day price
\begin{equation}
y^{(t)} = p_2^{(t)}.
\end{equation}

This setup is important.
Predictions are made \emph{after} the opening price of day $t$ is observed but \emph{before} day $t$ high, low, or close are realized.
That makes the close-direction problem a same-day forecasting task rather than a pure next-day task.
No day-$t$ high, low, or close value is used as an input feature for the corresponding day-$t$ prediction.

\subsection{Timing and Leakage Controls}

The timing convention is central to the interpretation of the results.
Because the day-$t$ opening price is included in the feature vector, the forecast is an after-open same-day forecast.
This is statistically valid for studying open-to-close directional predictability, but it is not equivalent to a forecast made before the trading day begins.

The benchmark therefore avoids three common sources of overstatement.
First, the walk-forward loop trains only on observations available before the target day.
Second, threshold-conditioned accuracy is reported as descriptive evidence with the corresponding number of observations, not as an optimized trading rule.
Third, the high and low targets are treated as statistical targets rather than directly executable trading signals, because the realized intraday path is not modeled.

\subsection{Models}

We benchmark the following models:
\begin{itemize}[noitemsep]
    \item \textbf{XGBoost} \citep{chen2016xgboost}: the primary tree-ensemble benchmark.
    \item \textbf{Random Forest} \citep{breiman2001random}: a strong bagging baseline for low-dimensional tabular data.
    \item \textbf{LightGBM} \citep{ke2017lightgbm}: a second boosting baseline.
    \item \textbf{Logistic Regression}: trained directly on direction labels rather than on prices.
    \item \textbf{Naive Momentum}: predicts today's direction equals yesterday's direction.
    \item \textbf{Naive Always-Up}: always predicts an increase.
    \item \textbf{Random}: predicts direction uniformly at random.
\end{itemize}

The tree models forecast continuous prices; direction is then derived by comparing $\hat{y}_t$ with $y_{t-1}$.
Logistic Regression predicts direction directly and therefore serves as a useful benchmark for the value of direct classification versus regression-derived direction.

\subsection{Walk-Forward Validation}

We use expanding-window walk-forward validation \citep{tashman2000out}.
The last 800 trading days form the main test period.
At each step, the model is trained on all data available up to day $t-1$, predicts day $t$, and then incorporates day $t$ into the training history for the next iteration.
This is more realistic than random shuffling or ordinary $k$-fold cross-validation, which would violate the temporal structure of the problem \citep{bergmeir2012use}.
The reported benchmark uses fixed model specifications rather than selecting hyperparameters from the final test sequence.

\subsection{Metrics}

For price-level prediction we report mean absolute percentage error (MAPE).
For directional prediction we report accuracy, precision, recall, and F1 where appropriate.
A prediction is directionally correct when
\begin{equation}
\mathbb{1}\big[(\hat{y}_t \ge y_{t-1}) = (y_t \ge y_{t-1})\big] = 1.
\end{equation}

We also evaluate \emph{threshold-conditioned} directional accuracy.
Let
\begin{equation}
\delta_t = \frac{|\hat{y}_t - y_{t-1}|}{y_{t-1}} \times 100
\end{equation}
denote the predicted absolute move in percent.
For each threshold $\tau$, we compute directional accuracy only on observations with $\delta_t \ge \tau$ and report the number of qualifying observations.
This is central to the paper because apparent high-accuracy regions can be driven by very small sample sizes.

\subsection{Statistical Tests}

We use three statistical tools:
\begin{itemize}[noitemsep]
    \item \textbf{Diebold--Mariano} tests \citep{diebold1995comparing, west1996asymptotic} for forecast-error comparisons between tree models.
    \item \textbf{McNemar} tests for paired directional-classification comparisons.
    \item \textbf{Bootstrap confidence intervals} with 1,000 resamples for XGBoost close-direction accuracy.
\end{itemize}

\subsection{Auxiliary Stock-Universe Screen}

We also summarize an auxiliary stock-wide analysis generated with the same XGBoost open-plus-lagged setup.
In that screen, 541 U.S. equities are summarized by the area under their threshold-accuracy curves (AUC), where higher AUC indicates that threshold-conditioned directional accuracy remains high across a broader range of predicted-move cutoffs.

We use that screen descriptively only.
It is not the basis for the statistical tests in this paper, and it does not replace the main SPY benchmark.
Its role is narrower: it tells us whether SPY is a plausible case study for the close target and how unusual the SPY results are relative to the broader stock set.
Because SPY ranks unusually well for the close target in this screen, the main SPY results should be interpreted as a favorable case study rather than as an average-stock estimate.

\section{Results}
\label{sec:results}

\subsection{Main Close-Target Benchmark}

Table~\ref{tab:close_benchmark} reports the main SPY close-target benchmark.
Two findings stand out.
First, Random Forest is the strongest tree baseline on both regression and direction.
Second, Logistic Regression substantially outperforms the regression-derived directional baselines, suggesting that direct classification is a serious alternative for this task.

\begin{table}[htbp]
\centering
\caption{SPY close-target benchmark on the last 800 trading days. XGBoost accuracy is shown with a 95\% bootstrap confidence interval.}
\label{tab:close_benchmark}
\small
\resizebox{\linewidth}{!}{
\begin{tabular}{@{}lccccc@{}}
\toprule
Model & MAPE (\%) & Accuracy (\%) & Precision (\%) & Recall (\%) & F1 (\%) \\
\midrule
XGBoost             & 0.882 & 58.45 [54.94, 62.08] & 65.94 & 43.33 & 52.30 \\
Random Forest       & \textbf{0.785} & 61.20 & 66.77 & 52.14 & 58.56 \\
LightGBM            & 1.275 & 52.82 & 62.87 & 25.00 & 35.78 \\
Logistic Regression & --    & \textbf{71.09} & -- & -- & -- \\
Naive Momentum      & --    & 50.25 & -- & -- & -- \\
Naive Always-Up     & --    & 52.57 & -- & -- & -- \\
Random              & --    & 49.94 & -- & -- & -- \\
\bottomrule
\end{tabular}
}
\end{table}

Although the paper centers XGBoost because it is the core model in the stock pipeline, Table~\ref{tab:close_benchmark} argues against an overly XGBoost-centric interpretation.
For the SPY close target, the more honest summary is that tree ensembles produce modest but real directional signal, while direct classification performs best overall.

\subsection{Target Comparison on SPY}

We report three same-day targets: high, low, and close.
On SPY, XGBoost attains lower MAPE and higher directional accuracy for high and low than for close, while Random Forest again dominates the tree models:
\begin{itemize}[noitemsep]
    \item \textbf{Close}: XGBoost MAPE 0.882 and accuracy 58.45\%; Random Forest MAPE 0.785 and accuracy 61.20\%.
    \item \textbf{High}: XGBoost MAPE 0.589 and accuracy 63.95\%; Random Forest MAPE 0.499 and accuracy 69.46\%.
    \item \textbf{Low}: XGBoost MAPE 0.671 and accuracy 64.58\%; Random Forest MAPE 0.577 and accuracy 68.21\%.
\end{itemize}

These numbers show that the same-day high and low are easier to model than the close on SPY.
However, the auxiliary stock-universe screen later in this section shows that SPY is not representative for those targets, which is why the inferential emphasis of this paper remains on the close target.

\subsection{Threshold-Conditioned Accuracy}

\begin{figure}[htbp]
    \centering
    \includegraphics[width=0.85\linewidth]{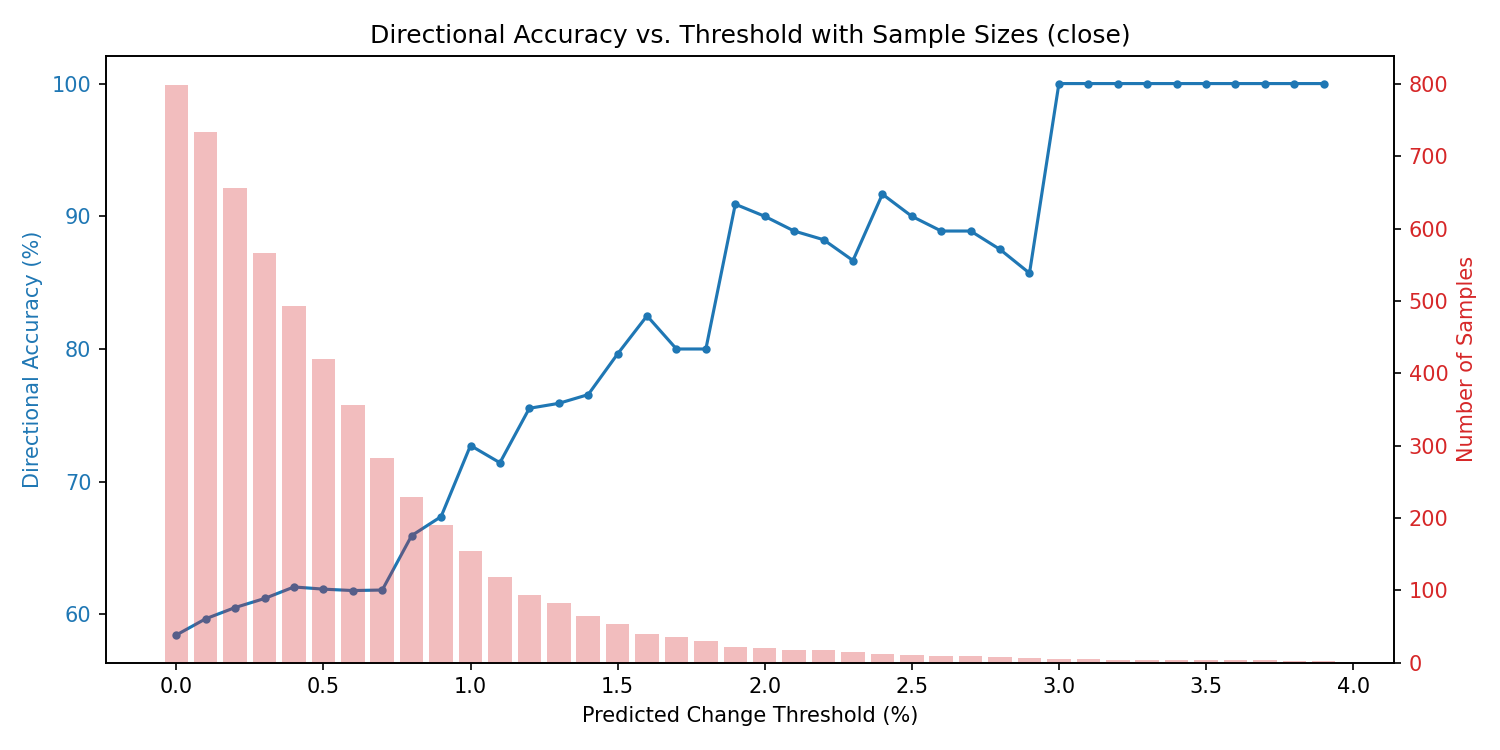}
    \caption{Threshold-conditioned close-direction accuracy for XGBoost on SPY. Accuracy increases with the predicted move threshold, but support decreases sharply.}
    \label{fig:threshold}
\end{figure}

Table~\ref{tab:threshold} and Figure~\ref{fig:threshold} summarize XGBoost close-direction accuracy as a function of predicted move size.
The qualitative pattern is intuitive: when the model predicts larger moves, directional accuracy improves.
The crucial counterweight is that the effective sample size collapses quickly.

\begin{table}[htbp]
\centering
\caption{XGBoost close-direction accuracy by predicted absolute move threshold on SPY.}
\label{tab:threshold}
\small
\begin{tabular}{@{}ccccccccc@{}}
\toprule
Threshold (\%) & 0.0 & 0.5 & 1.0 & 1.5 & 2.0 & 2.5 & 3.0 & 3.5 \\
\midrule
$N$            & 799 & 420 & 154 & 54 & 20 & 10 & 5 & 3 \\
Accuracy (\%)  & 58.4 & 61.9 & 72.7 & 79.6 & 90.0 & 90.0 & 100.0 & 100.0 \\
\bottomrule
\end{tabular}
\end{table}

This table is the clearest example of why sample-size disclosure matters.
The 90\%--100\% region is real in the sense that it appears in the held-out test set, but it is based on 20, 10, 5, and 3 observations respectively.
Those results are therefore descriptive rather than stable estimates of a deployable decision rule.

\subsection{Statistical Tests}

Table~\ref{tab:stats} compares XGBoost and Random Forest.
For the close target, Diebold--Mariano rejects equality of squared forecast errors in favor of Random Forest, while McNemar does not reject equal directional accuracy at the 5\% level.
This is consistent with the headline result: Random Forest is the better regressor, but the gap in close-direction classification is smaller than the regression-error gap alone would suggest.

\begin{table}[htbp]
\centering
\caption{XGBoost vs.\ Random Forest statistical tests on SPY.}
\label{tab:stats}
\small
\begin{tabular}{@{}llcc@{}}
\toprule
Target & Test & Statistic & $p$-value \\
\midrule
\multirow{2}{*}{Close}
& Diebold--Mariano & 3.657 & $< 0.001$ \\
& McNemar          & 3.150 & 0.076 \\
\midrule
\multirow{2}{*}{High}
& Diebold--Mariano & 3.739 & $< 0.001$ \\
& McNemar          & 16.219 & $< 0.001$ \\
\midrule
\multirow{2}{*}{Low}
& Diebold--Mariano & 4.126 & $< 0.001$ \\
& McNemar          & 6.588 & 0.010 \\
\bottomrule
\end{tabular}
\end{table}

\subsection{Regime-Specific Results}

To assess robustness, we evaluate XGBoost on three non-overlapping historical regimes.
The pattern is not one of complete stability, but it is also not random noise:
the COVID-era period is the easiest, the 2013--2019 bull market is the hardest, and all three periods remain above 55\% directional accuracy.

\begin{table}[htbp]
\centering
\caption{XGBoost close-target results on three historical subperiods.}
\label{tab:periods}
\small
\begin{tabular}{@{}lccc@{}}
\toprule
Period & $N_{\text{train}} / N_{\text{test}}$ & MAPE (\%) & Directional Accuracy (\%) \\
\midrule
2007--2012 (Crisis) & 1,056 / 452 & 0.827 & 60.98 \\
2013--2019 (Bull)   & 1,232 / 528 & 0.756 & 55.60 \\
2020--2023 (COVID)  & 629 / 269   & 0.836 & \textbf{65.30} \\
\bottomrule
\end{tabular}
\end{table}

This regime split is also consistent with separate training-window sensitivity analyses.
The strongest period-level result does not come from the longest history, which is consistent with the idea that very old data can introduce obsolete regimes rather than useful signal.

\subsection{Stock-Universe Context}

Table~\ref{tab:screen_summary} summarizes the auxiliary 541-equity screen.
The screen is useful precisely because it complicates the story.
It shows that threshold-conditioned predictability depends strongly on both the target and the asset.

\begin{table}[htbp]
\centering
\caption{Descriptive summary of the auxiliary 541-equity XGBoost screen. AUC refers to the area under the threshold-accuracy curve.}
\label{tab:screen_summary}
\small
\begin{tabular}{@{}lcccc@{}}
\toprule
Target & Median AUC & IQR AUC & Stocks with AUC $> 400$ & SPY Rank \\
\midrule
Low   & 435.46 & [419.56, 446.07] & 490 / 541 & 513 / 541 \\
High  & 431.97 & [415.86, 444.92] & 478 / 541 & 540 / 541 \\
Close & 350.50 & [330.36, 367.58] & 11 / 541  & \textbf{4 / 541} \\
\bottomrule
\end{tabular}
\end{table}

Two implications follow.
First, the close target is genuinely harder across the stock universe than high and low under this metric.
Second, SPY is unusually strong for the close target but unusually weak for high and low.
That asymmetry motivates the present paper's design: a close-target benchmark on SPY is defensible, while broad claims about SPY high/low predictability would be much weaker.

For completeness, the top five close-target AUC names in the screen are JNJ (420.09), WY (417.78), PPG (415.67), SPY (415.24), and DD (415.10).
This makes SPY a favorable close-target case in the broader equity screen.
The appropriate interpretation is therefore case-study evidence about a liquid index ETF, not a claim that the same level of close-direction predictability is typical across equities.

\subsection{Feature Importance}

\begin{figure}[htbp]
    \centering
    \includegraphics[width=0.85\linewidth]{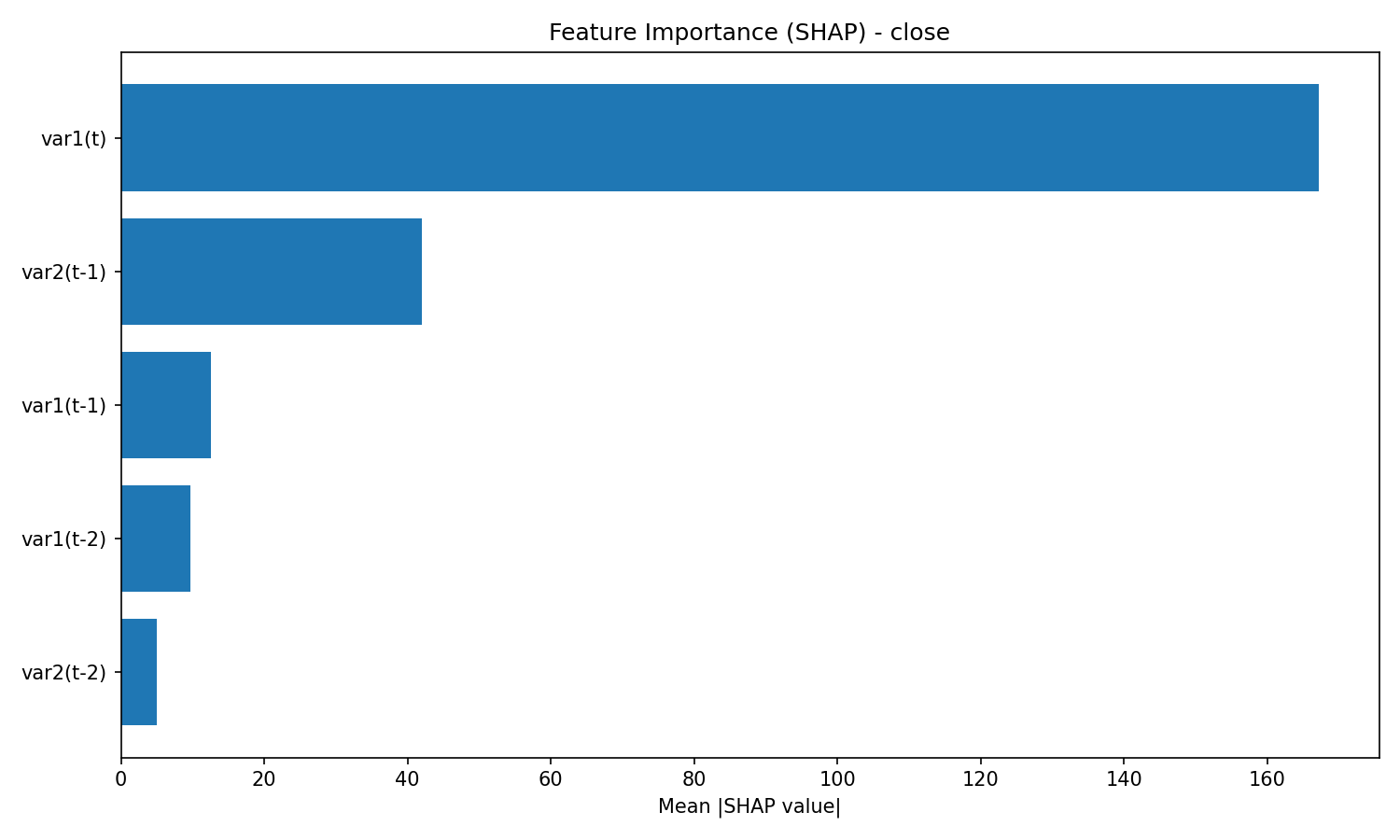}
    \caption{Mean absolute SHAP values for the XGBoost SPY close-target model.}
    \label{fig:shap}
\end{figure}

Figure~\ref{fig:shap} reports SHAP values for the XGBoost close-target model.
The picture is simple and plausible: today's open and the most recent lagged prices dominate older lags, which is what we would expect in a same-day forecasting problem.

\section{Discussion}
\label{sec:discussion}

\subsection{What the Results Support}

The strongest claim supported by the evidence is modest: simple open-plus-lagged daily features contain statistically detectable information about same-day SPY direction.
That claim is backed by walk-forward validation, bootstrap intervals, and cross-model comparisons.
It is \emph{not} a claim that XGBoost is the universally best model, nor that these results alone imply a production-ready trading strategy.

In fact, one of the most useful results is that Logistic Regression outperforms the regression-derived direction benchmarks on the close target.
This implies that if the objective is close-direction prediction alone, direct classification may be preferable to predicting prices first and extracting sign afterwards.
XGBoost remains valuable because it yields continuous forecasts, supports threshold filtering, and offers interpretable feature attributions.

\subsection{Why Economic Claims Remain Limited}

The statistical benchmark is stronger than any economic claim that could be made from the present setup alone.
A realistic trading study would require more careful treatment of execution timing, slippage, and market impact.
For that reason, the paper intentionally avoids claiming that the present evidence by itself establishes a deployable strategy.

\subsection{Limitations}

Several limitations remain:
\begin{enumerate}[noitemsep]
    \item \textbf{Single-instrument inference}: the main statistical tests are performed on SPY close only.
    \item \textbf{Descriptive stock screen}: the 541-equity screen is descriptive context, not a uniformly controlled replication study.
    \item \textbf{Favorable-case selection}: SPY ranks highly for close-target AUC in the auxiliary screen, so the main case study may overstate average-stock predictability.
    \item \textbf{Target asymmetry}: SPY behaves very differently across close versus high/low in the auxiliary screen, so target choice materially affects conclusions.
    \item \textbf{Small support at high thresholds}: the most impressive threshold-conditioned accuracies are based on very few observations.
    \item \textbf{Model misspecification}: the benchmark does not exhaust the design space; direct classifiers, better calibration, or alternative features may improve results.
    \item \textbf{Non-stationarity}: all financial prediction studies face structural breaks and time-varying market efficiency \citep{lo2004adaptive, timmermann2004efficient}.
\end{enumerate}

\section{Reproducibility and Scope}
\label{sec:reproducibility}

The manuscript and supporting material are archived at \url{https://osf.io/6thqk}.
The empirical results in this paper use the stock-training path only.
The main SPY benchmark is reproduced by the accompanying experiment script.
The auxiliary stock-universe screen is summarized by the accompanying stock-screen summary script.
Cryptocurrency data, cryptocurrency-specific feature pipelines, and automated trading utilities in the wider repository are not used for any result reported here.

The reproducibility target is intentionally narrow: reproduce the reported statistical forecasts, threshold-conditioned accuracy tables, regime summaries, statistical tests, and SHAP feature-importance figures for the stock experiments.
The paper does not present or rely on a live trading system.

\section{Conclusion}
\label{sec:conclusion}

This paper presents a conservative walk-forward benchmark for directional stock prediction from daily OHLC data.
On SPY, simple models trained at the open can recover modest but statistically meaningful same-day directional signal, especially when the model predicts larger moves.
Random Forest is the best tree-based baseline in our tests, and Logistic Regression is the strongest overall close-direction model.

The broader equity-screen context also matters.
An auxiliary 541-equity screen shows that target-specific predictability varies widely across assets, and that SPY is unusually strong for the close target but not for high or low.
That is precisely why the paper is framed as a close-target SPY benchmark with limited generalization claims.

Future work should focus on three extensions:
\begin{enumerate}[noitemsep]
    \item direct binary classification or calibrated probability forecasting,
    \item more carefully standardized cross-sectional replications across equities, and
    \item execution-aware economic evaluation that respects the timing assumptions of the forecasting setup.
\end{enumerate}

Within those limits, the main result is still useful: daily stock direction is hard, but not entirely featureless, and a transparent statistical benchmark is a more durable contribution than an overstated trading narrative.

\input{appendix}

\bibliography{references}

\end{document}

%% file: appendix.tex
\clearpage
\appendix
\renewcommand{\thesection}{Appendix \Alph{section}}
\section*{Appendices}
\addcontentsline{toc}{section}{Appendices}

\section{Supplementary Method Details}
\label{app:method}

\subsection*{A.1 Walk-Forward Prediction Algorithm}

Algorithm~\ref{alg:walkforward} summarizes the expanding-window validation procedure used for the stock experiments.
At each test step, the model is fit only on observations available before the prediction day.

\begin{algorithm}[H]
\caption{Expanding-Window Walk-Forward Prediction}
\label{alg:walkforward}
\begin{algorithmic}[1]
\Require Initial training data $\mathcal{D}_{\text{train}}$
\Require Ordered test data $\mathcal{D}_{\text{test}} = \{(\mathbf{x}_1, y_1), \ldots, (\mathbf{x}_T, y_T)\}$
\Ensure One-step-ahead predictions $\hat{y}_1, \ldots, \hat{y}_T$
\State $\mathcal{H} \leftarrow \mathcal{D}_{\text{train}}$
\For{$t = 1$ to $T$}
    \State Fit forecasting model $f_t$ on $\mathcal{H}$
    \State $\hat{y}_t \leftarrow f_t(\mathbf{x}_t)$
    \State $\mathcal{H} \leftarrow \mathcal{H} \cup \{(\mathbf{x}_t, y_t)\}$
\EndFor
\end{algorithmic}
\end{algorithm}

\subsection*{A.2 XGBoost Objective and Implementation}

XGBoost constructs an additive ensemble of $K$ regression trees,
\begin{equation}
\hat{y}_i = \sum_{k=1}^{K} f_k(\mathbf{x}_i), \quad f_k \in \mathcal{F},
\end{equation}
where $\mathcal{F}$ is the space of regression trees.
The regularized training objective is
\begin{equation}
\mathcal{L} = \sum_{i=1}^{n} l(y_i, \hat{y}_i) + \sum_{k=1}^{K} \Omega(f_k),
\end{equation}
with regularization
\begin{equation}
\Omega(f) = \gamma T + \frac{1}{2}\lambda \sum_{j=1}^{T} w_j^2,
\end{equation}
where $T$ is the number of leaves and $w_j$ are leaf weights.
The stock benchmark uses squared error loss and the lightweight XGBoost configuration used in the repository's stock pipeline.
The supplementary experiment script also evaluated a tuned close-target configuration with 100 estimators, learning rate 0.05, maximum depth 6, minimum child weight 4, subsampling 0.7, and column subsampling 0.7.
That tuned configuration did not improve close-direction accuracy relative to the basic XGBoost specification.

\subsection*{A.3 Additional Metrics}

For price-level prediction, the supplementary result files include mean squared error (MSE), root mean squared error (RMSE), mean absolute error (MAE), mean absolute percentage error (MAPE), and $R^2$:
\begin{align}
\text{MAPE} &= \frac{100\%}{n}\sum_{t=1}^{n}\left|\frac{y_t - \hat{y}_t}{y_t}\right|, \\
R^2 &= 1 - \frac{\sum_{t=1}^{n}(y_t - \hat{y}_t)^2}{\sum_{t=1}^{n}(y_t - \bar{y})^2}.
\end{align}

\section{Supplementary Results}
\label{app:results}

\subsection*{B.1 Regression Metrics Across Targets}

Table~\ref{tab:appendix_regression} reports supplementary price-level regression metrics for the tree models.
These results support the main text's statement that Random Forest is the strongest price-level regressor in this benchmark.

\begin{table}[H]
\centering
\caption{Supplementary regression performance on SPY using walk-forward validation on the last 800 trading days.}
\label{tab:appendix_regression}
\small
\resizebox{\linewidth}{!}{
\begin{tabular}{@{}llccccc@{}}
\toprule
Target & Model & MSE & RMSE & MAE & MAPE (\%) & $R^2$ \\
\midrule
\multirow{4}{*}{Close}
& XGBoost       & 21.882 & 4.678 & 3.737 & 0.882 & 0.980 \\
& XGBoost tuned & 45.881 & 6.774 & 5.661 & 1.318 & 0.957 \\
& Random Forest & \textbf{17.568} & \textbf{4.191} & \textbf{3.307} & \textbf{0.785} & \textbf{0.984} \\
& LightGBM      & 47.450 & 6.888 & 5.483 & 1.275 & 0.956 \\
\midrule
\multirow{3}{*}{High}
& XGBoost       & 10.611 & 3.257 & 2.506 & 0.589 & 0.990 \\
& Random Forest & \textbf{7.608} & \textbf{2.758} & \textbf{2.124} & \textbf{0.499} & \textbf{0.993} \\
& LightGBM      & 36.788 & 6.065 & 4.504 & 1.037 & 0.965 \\
\midrule
\multirow{3}{*}{Low}
& XGBoost       & 13.184 & 3.631 & 2.833 & 0.671 & 0.988 \\
& Random Forest & \textbf{9.916} & \textbf{3.149} & \textbf{2.428} & \textbf{0.577} & \textbf{0.991} \\
& LightGBM      & 38.912 & 6.238 & 4.799 & 1.119 & 0.965 \\
\bottomrule
\end{tabular}
}
\end{table}

\subsection*{B.2 Direct Classification Across Targets}

The direct Logistic Regression baseline is included to separate price-level prediction from directional classification.
In the supplementary results, Logistic Regression reaches 71.09\% directional accuracy for close, 81.10\% for high, and 80.10\% for low.
These numbers reinforce the main conclusion that direct binary classification is a promising extension for future work, although the high and low targets should not be interpreted as directly executable trading signals without an intraday execution model.

\subsection*{B.3 Execution-Illustrative Backtest}

Earlier repository drafts included a simple long/flat backtest based on XGBoost close-direction signals.
Because the same-day forecast uses the day-$t$ open and evaluates close-to-close changes, the backtest is retained here only as an execution-illustrative appendix rather than as primary evidence.
It should not be read as a production trading strategy.

\begin{table}[H]
\centering
\caption{Illustrative long/flat backtest using XGBoost close-direction signals. B\&H denotes buy-and-hold.}
\label{tab:appendix_backtest}
\small
\resizebox{\linewidth}{!}{
\begin{tabular}{@{}lccccc@{}}
\toprule
Cost (bps) & Return (\%) & B\&H Return (\%) & Sharpe & Max DD (\%) & Trades \\
\midrule
0   & 239.36 & 34.63 & 3.456 & 5.76 & 300 \\
5   & 192.17 & 34.63 & 3.043 & 6.45 & 300 \\
10  & 151.52 & 34.63 & 2.626 & 7.16 & 300 \\
20  & 86.36  & 34.63 & 1.785 & 8.89 & 300 \\
\bottomrule
\end{tabular}
}
\end{table}

\begin{figure}[H]
    \centering
    \includegraphics[width=0.85\linewidth]{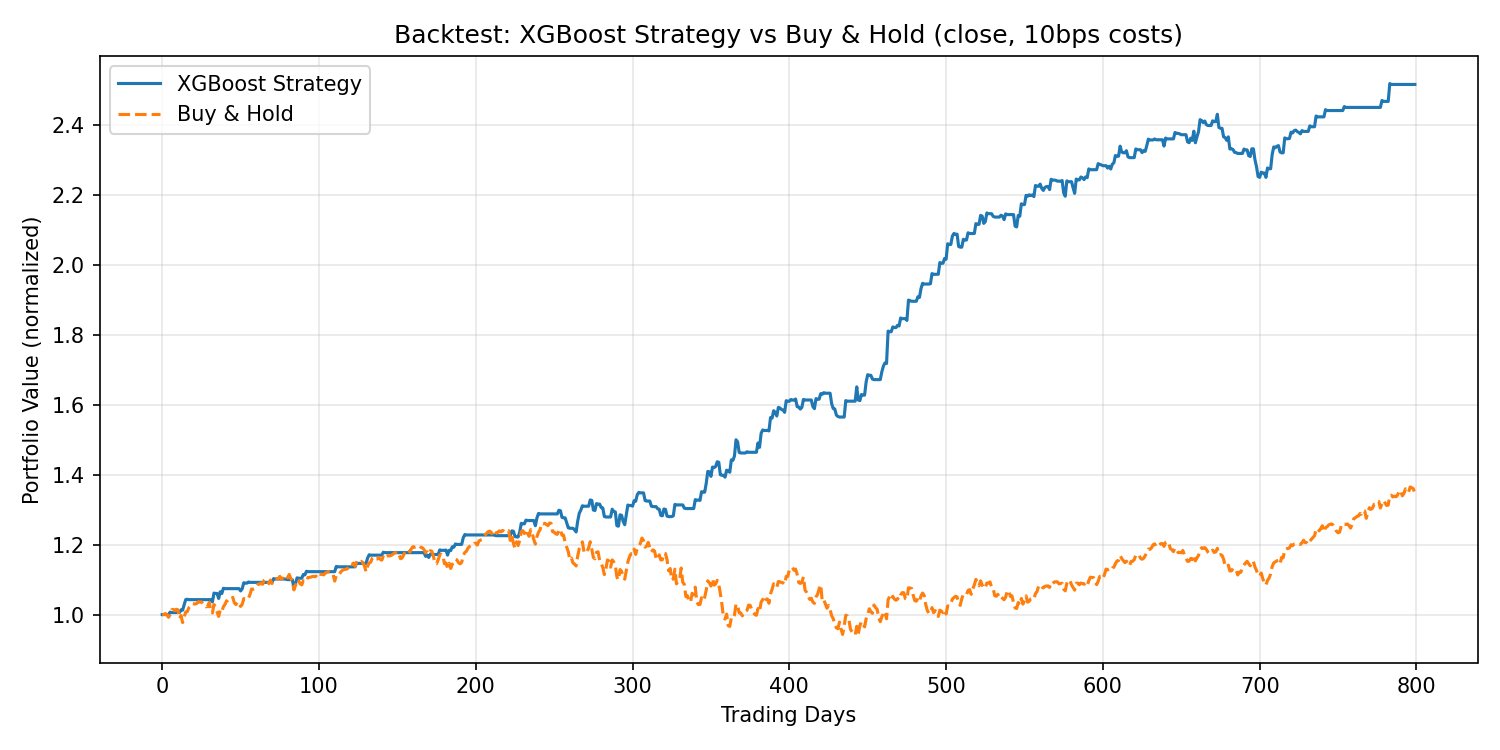}
    \caption{Illustrative cumulative portfolio value for the XGBoost close-direction long/flat rule with 10 bps costs versus buy-and-hold.}
    \label{fig:appendix_backtest}
\end{figure}

\subsection*{B.4 Training-Window Sensitivity}

The regime-specific results also provide indirect evidence about training-window sensitivity.
The COVID-era model, trained on 629 observations, attains the highest close-direction accuracy among the three reported regimes, while the longer bull-market window is weaker.
Because market regimes differ, this should not be interpreted as a causal estimate of the optimal training length.
It does suggest, however, that using more distant historical observations does not automatically improve short-horizon directional forecasting in non-stationary equity data.

\subsection*{B.5 Supplementary Target Figures}

\begin{figure}[H]
    \centering
    \includegraphics[width=0.85\linewidth]{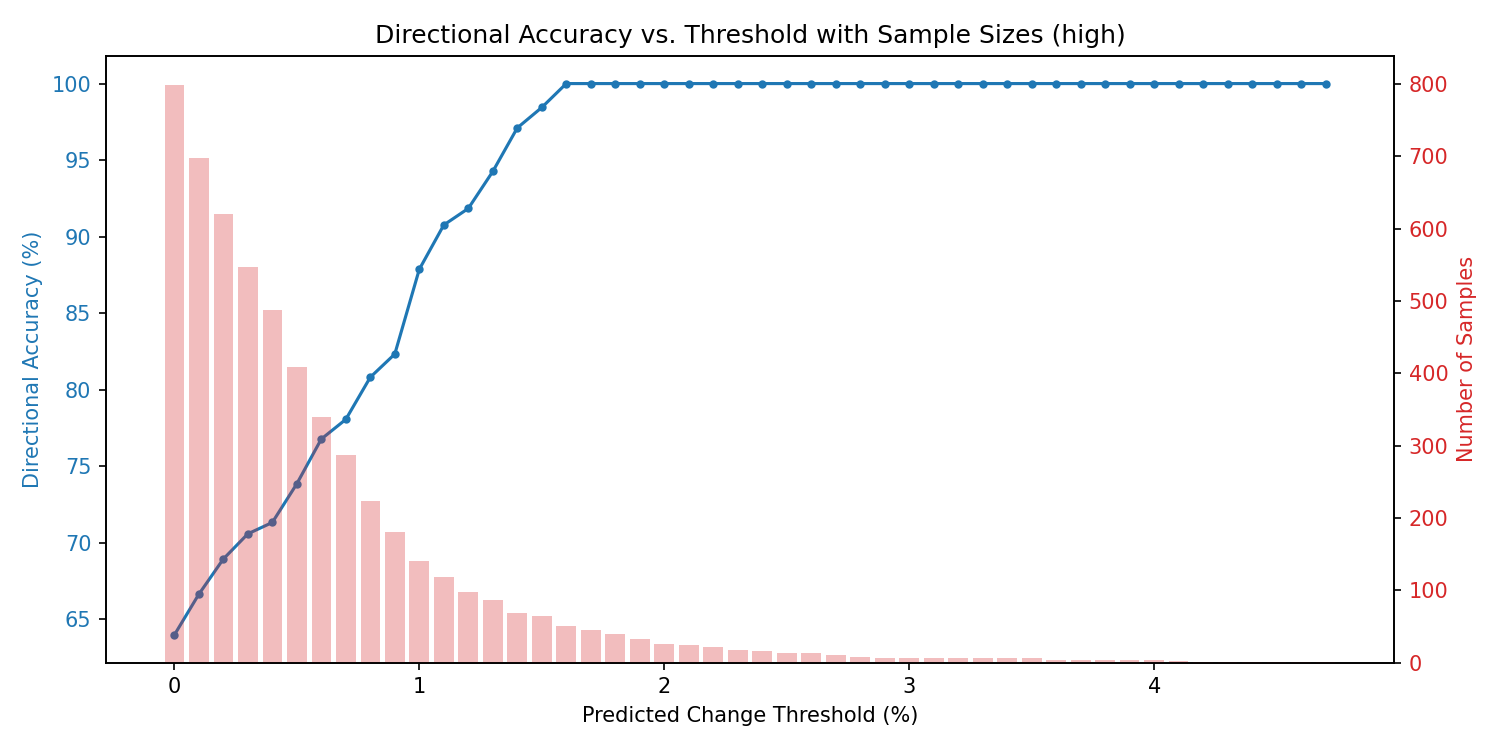}
    \caption{Threshold-conditioned directional accuracy for the XGBoost SPY high-target model.}
    \label{fig:appendix_threshold_high}
\end{figure}

\begin{figure}[H]
    \centering
    \includegraphics[width=0.85\linewidth]{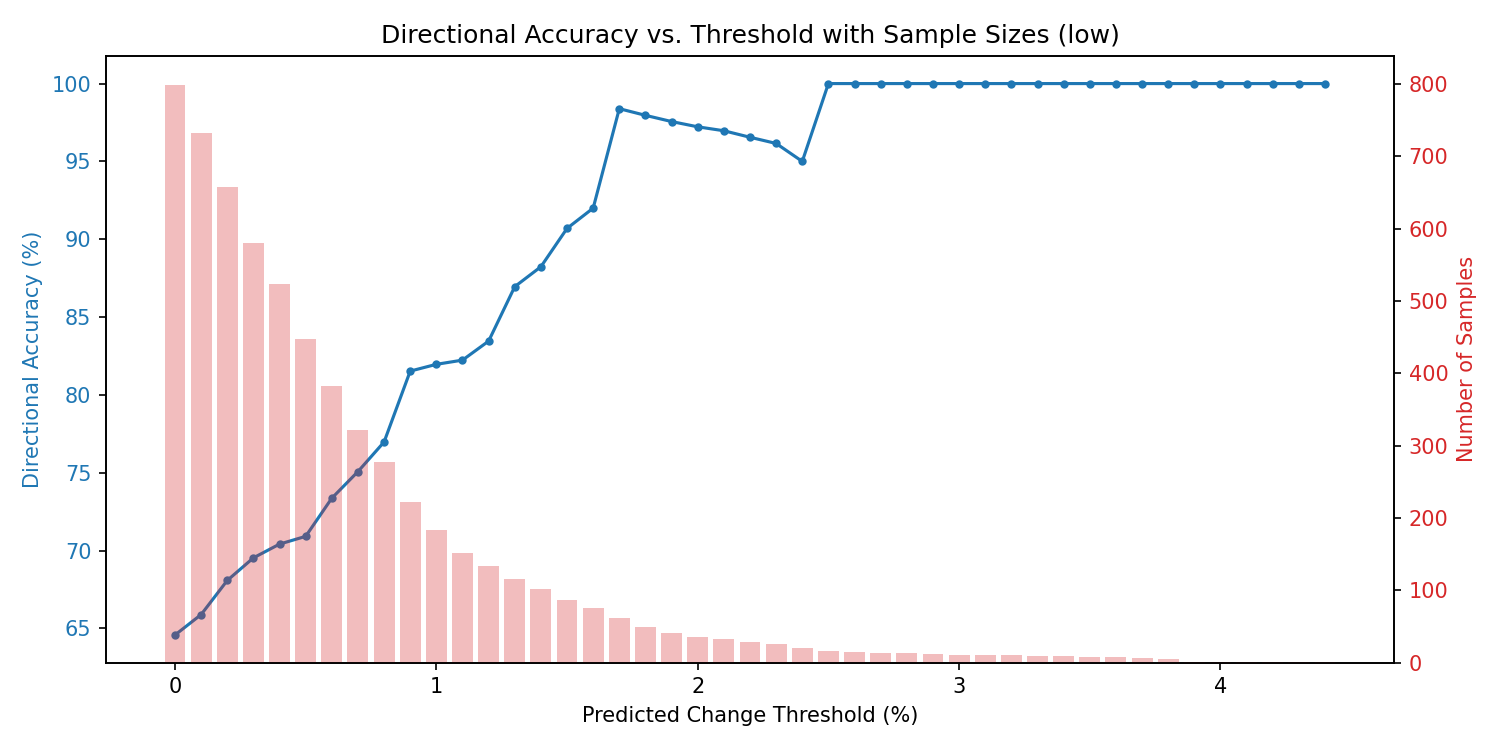}
    \caption{Threshold-conditioned directional accuracy for the XGBoost SPY low-target model.}
    \label{fig:appendix_threshold_low}
\end{figure}

\begin{figure}[H]
    \centering
    \includegraphics[width=0.85\linewidth]{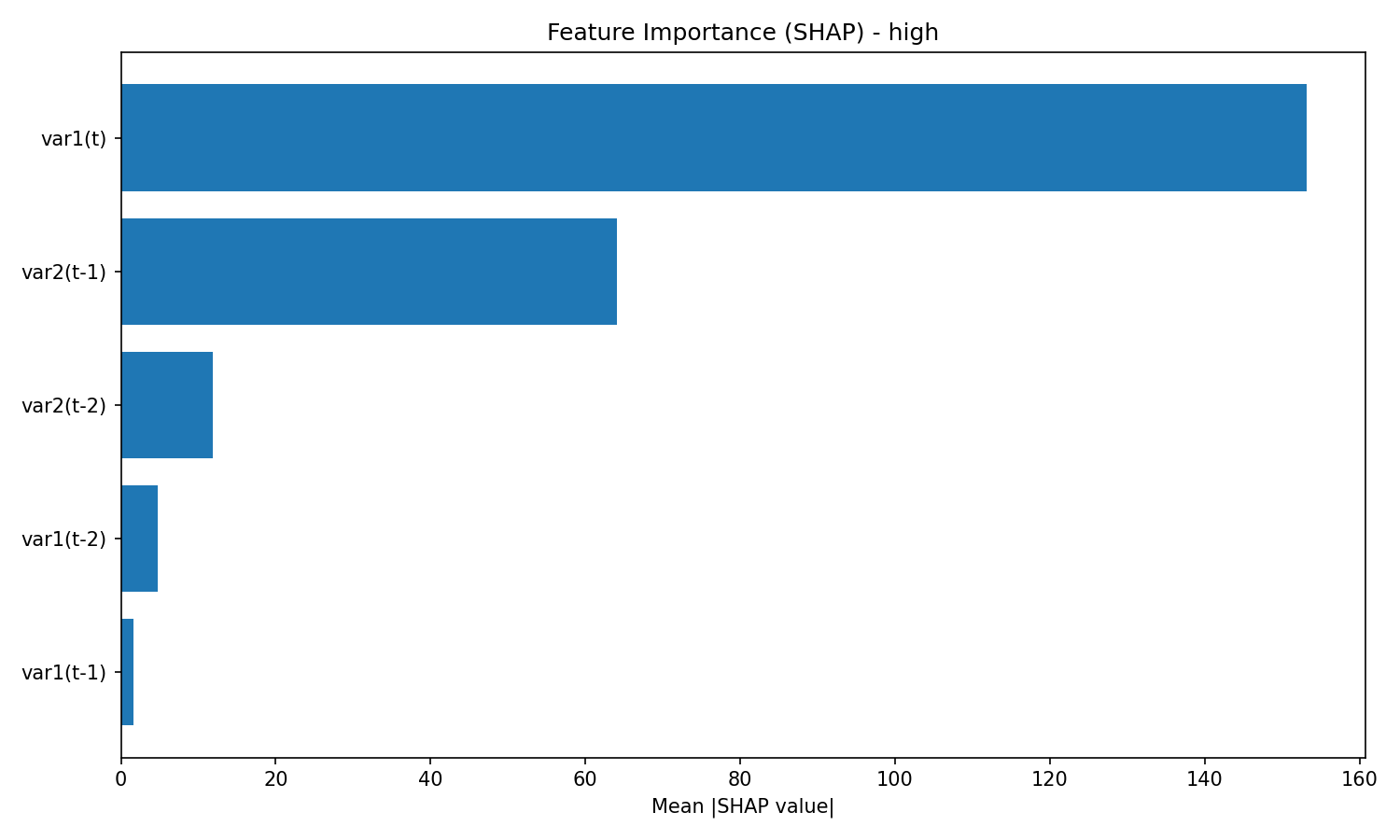}
    \caption{Mean absolute SHAP values for the XGBoost SPY high-target model.}
    \label{fig:appendix_shap_high}
\end{figure}

\begin{figure}[H]
    \centering
    \includegraphics[width=0.85\linewidth]{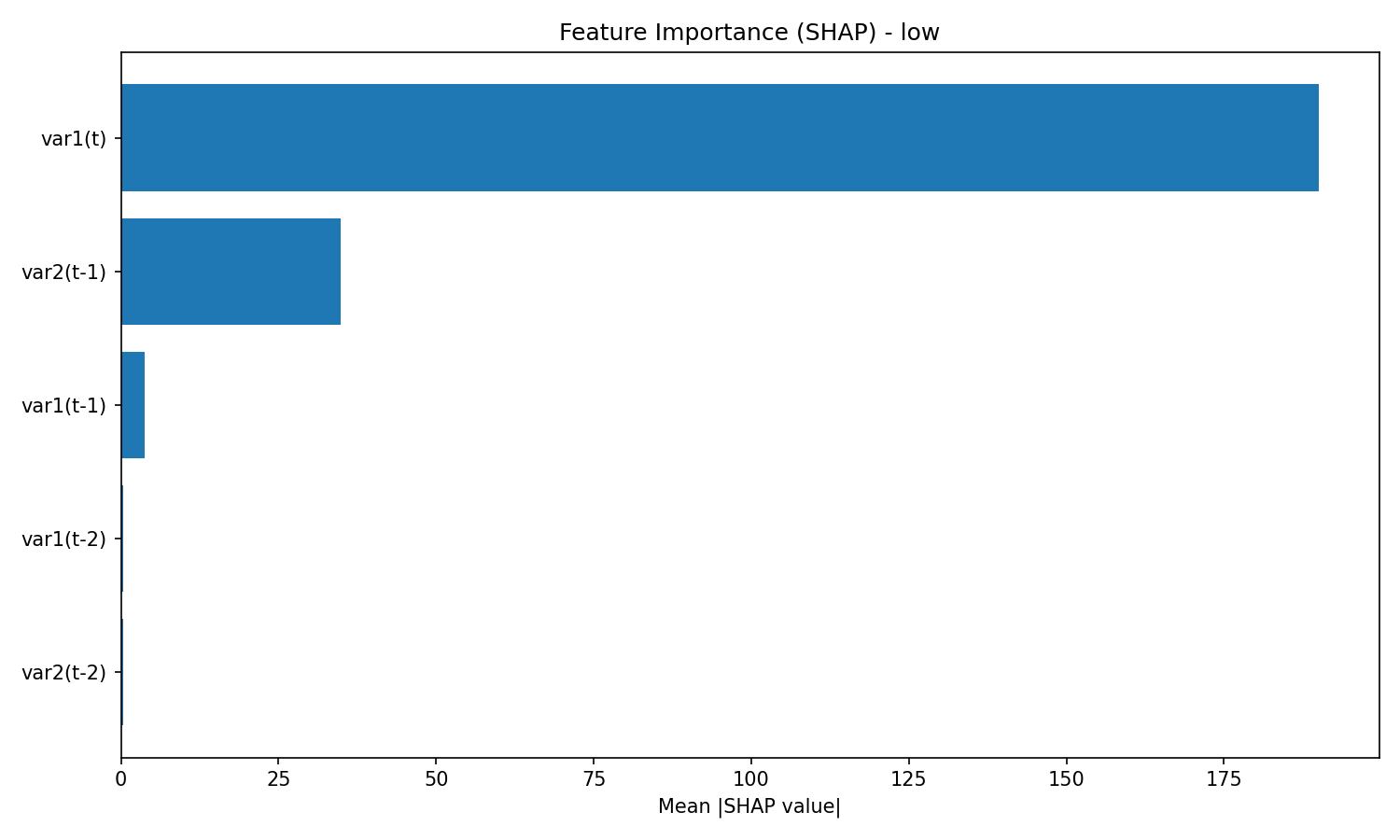}
    \caption{Mean absolute SHAP values for the XGBoost SPY low-target model.}
    \label{fig:appendix_shap_low}
\end{figure}

Figures~\ref{fig:appendix_threshold_high}--\ref{fig:appendix_shap_low} provide supplementary diagnostics for the high and low targets.
These figures are included for completeness because the same stock-training pipeline evaluates high, low, and close targets.
As discussed in the main text, the high and low targets should be interpreted as statistical forecasting targets rather than directly executable trading signals.

